\documentstyle[here,epsfig]{mn}
\begin{document}
\def\simlt{\mathrel{\rlap{\lower 3pt\hbox{$\sim$}}\raise 2.0pt\hbox{$<$}}}
\def\simgt{\mathrel{\rlap{\lower 3pt\hbox{$\sim$}} \raise
2.0pt\hbox{$>$}}}

\def\whs{\thinspace ${\rm W \: Hz^{-1} \: sr^{-1}}$}
\def\lpow{\thinspace ${\rm log_{10}P_{1.4GHz}\: }$}

\title[] {Faint radio--loud quasars: clues to their evolution} 
\author[M. Cirasuolo,
M. Magliocchetti, A. Celotti] {M. Cirasuolo, M. Magliocchetti,
A. Celotti \\ S.I.S.S.A., Via Beirut 2-4, I-34014 Trieste, Italy } \maketitle
\vspace {7cm }

\begin{abstract}
The quasar sample selected by cross-correlating the FIRST and the 2dF
Quasar Redshift Surveys allows us to explore, for the first time, the
faint end of the radio and optical luminosity functions up to $z \simeq
2.2$. We find indications ($\sim 3 \sigma$) of a negative evolution
for these faint sources at $z \simgt 1.8$, both in radio and optical bands.
This  corresponds to a decrement in the space density of faint quasars
of about a factor 2 at $z$=2.2 and confirms the presence of a
differential evolution for the population of radio-active quasars.
The faint end of both luminosity functions flattens and the comparison
with the (optical) number density of the whole quasar population
supports a dependence of the fraction of radio detected quasars on the
optical luminosity.  A progressive decrease in the fraction
of quasars in the whole radio source population can be consistently
accounted for within the `receding torus' scenario. The population of
low luminosity quasars, which the FIRST-2dF detects,
appears to depart from the `classical' scheme for radio-loud quasars.

\end{abstract}
\begin{keywords} galaxies: active - cosmology: observations - 
radio continuum: quasars
\end{keywords}
%
\section{INTRODUCTION}
It is becoming increasingly evident that the formation of
super-massive black holes (BH) powering nuclear activity is intimately
related to the formation of their host galaxies.  Observational
evidence such as the presence of massive BH ($10^6 - 10^9 M_{\odot}$ )
in at least all local galaxies with a spheroidal component (Kormendy
\& Richstone 1995) and tight correlations between the BH mass and the
overall properties of the spheroids supports this connection
(Magorrian et al. 1998; McLure \& Dunlop 2002; Ferrarese 2002).  The
evolution of nuclear activity (AGN) with cosmic time is then a key
ingredient in order to constrain the formation of galactic systems.
Investigations of radio-emitting AGN can be of great help in tackling
such issue since -- as already proved by very early studies (Longair
1966) -- radio emission can be used as a clean tool, not affected by
obscuration, to trace the AGN evolution. On the other hand, the
connection between the radio and optical activity still needs to be
understood.

One of the key studies on the evolution of both radio galaxies and
radio quasars, was provided by Dunlop \& Peacock (1990).  The radio
luminosity function (LF) of these populations was separately obtained
for steep spectrum and flat spectrum sources, by means of several
samples selected at 2.7 GHz with relatively bright flux limits
(ranging from 0.1 to 2.0 Jy). This analysis confirmed the presence of
a so-called `redshift cut-off'' for the population of flat spectrum
sources (Peacock 1985), with a decline in comoving density of a factor
$\sim 5$ between $z =2$ and $z=4$.  Most intriguingly, their work
provided the first evidence for a similar behaviour for the population
of steep spectrum sources, both quasars and radio galaxies, making
this decline a common feature for all powerful radio objects. The
evolution of both populations was found to be satisfactorily described
by a pure luminosity evolution (PLE) model, analogous to that found
for optically selected quasars (Boyle et al. 1988; Boyle et al. 2000;
Croom et al. 2004), as well as by a luminosity/density evolution
model, which incorporates a negative density evolution at high
redshifts.

In order to investigate the possible presence of a redshift cut-off in
the AGN evolution, several searches for high redshift radio--loud
quasars have been carried out in the last years.  Shaver et al. (1996,
1999) argued for a drop in space density of flat spectrum quasars by
more than a factor 10 between $z\sim 2.5$ and $z\sim 6$. 
However, a re-analysis of such sources (Jarvis \& Rawlings 2000;
Jarvis et al. 2001) suggests a more gradual (factor $\sim 4$) decline
in the same redshift interval.  A luminosity dependent cut-off, with
a space density decline less dramatic for the most luminous radio
sources, has also been claimed by Dunlop (1998) from a recent update
of the work by Dunlop \& Peacock (1990).

The question of a redshift cut-off in the quasar evolution has also
been addressed at other wavelengths.  No evidence for a rapid decline
of the quasar space density has been found for soft X-ray selected
samples at $z \simgt 2.5$ (Miyaji, Hasinger \& Schmidt 2000, Ueda et
al. 2003).  On the other hand, optical data from the Sloan survey (Fan
et al. 2001) suggest the space density of ${\rm M_{AB}(1450 \AA) <
-25.5}$ quasars at $z \sim 4$ to be more than one order of magnitude
lower than that found at $z \sim 2$ (Boyle et al. 2000).  Recently, by
using a sample of 13 optically luminous radio quasars, Vigotti et
al. (2003) showed that their decline in the space density is about a
factor of 2 between $z\sim 2$ and $z \sim 4$, significantly smaller
than the value $\sim 10$ found for samples including lower luminosity
objects (Fan et al. 2001).

However, the comparison with the behaviour of radio--selected sources
hinges on the connection between the optical (X--ray) and radio
activity. This is clearly of paramount relevance also with regards to
the physical connection between the accretion processes and the jet
formation. This aspect can be tackled from the comparison of the
radio--loud and radio--quiet quasar populations.

The behaviour of radio sources is also tightly connected within the
framework of unification models (e.g. Padovani \& Urry 1992, Urry \&
Padovani 1995, Jackson and Wall 1999).  These rely on the fact that
beaming effects and thus orientation play a key role in determining
the observed properties of sources: BL Lac objects and flat spectrum
quasars would be the beamed version -- i.e. viewed at an angle close
to the jet axis -- of FRI and FRII radio galaxies (Fanaroff \& Riley
1974), respectively. The behaviour of radio quasars is thus closely
related to that of the whole radio source population and in particular
of powerful radio galaxies.

In the light of the above discussion the aim of the present paper is
to estimate the low luminosity end of the LF of radio active quasars,
by taking advantage of the new final releases of the 2dF QSO Redshift
and the FIRST surveys.  The aims are: i) to explore the cosmological
evolution of the selected sources, ii) to examine the relation between
the radio and optical activity and iii) the connection between quasars
and radio galaxies.

The layout of the paper is as follows. In Section 2 we give a brief
description of the new final releases of FIRST and 2dF Quasar Redshift
Survey datasets used for our analysis and the matching procedure used
to cross-correlate the catalogues. In Section 3 we describe the
computation of the radio and optical luminosity functions and present
our results, giving an assessment for incompleteness effects.  Our
findings are compared with other LFs already present in literature in
Section 4.  We discuss the results and summarize our conclusions in
Section 5.  For sake of comparison with previous works we will
adopt $H_0 = 50 \; {\rm km \; s^{-1} Mpc^{-1}}$, $q_0=0.5 $ and
$\Lambda = 0$ (hereafter cosmology I), but also report our results on
both the radio and optical LFs within the ``concordance'' model,
consistent with the Wilkinson Microwave Anisotropy Probe data (Bennett
et al. 2003), i.e.: $\Omega_{\rm M}=0.3$, $\Omega_\Lambda=0.7$ and
$H_0=70~{\rm km~s^{-1}}$ (cosmology II).

\section{THE DATA SETS}

\subsection{The FIRST Survey}\label{FIRST}
The FIRST (Faint Images of the Radio Sky at Twenty centimeters) survey
(Becker et al. 1995) began in 1993 and its latest release (April
2003), used for this work, contains 811,117 sources observed at {\rm
1.4 GHz} down to a flux limit ${\rm S_{1.4GHz} \simeq 0.8}$ mJy.  The
survey covers a total of about 9033 square degrees of sky (8422 square
degrees in the north Galactic cap and 611 in the south Galactic cap)
and is substantially complete: it has been estimated to be 95 and 80
per cent complete at 2 mJy and 1 mJy, respectively (Becker et
al. 1995). Note that, as the completeness level quickly drops for flux
levels fainter than 2 mJy, in the following analysis we will only
consider sources brighter than this limit.

The surface density of objects in the catalogue is $\sim 90$ per
square degree, though this is reduced to $\sim 80$ deg$^{-2}$ if one
combines multi-component sources (Magliocchetti et al. 1998).  The
accuracy of the positions depends both on the brightness and size of
the sources and on the noise in the map. Point sources at the
detection limit of the catalogue have positions accurate to better
than 1 arcsec (90 per cent confidence level); 2 mJy point sources in
typically noisy regions have positions determined to better than 0.5
arcsec.

\subsection{The 2dF QSO Redshift Survey}
The 2dF QSO Redshift Survey (2QZ) has recently been completed and in
this work we will refer to the final release (Croom et al. 2004).
Here we briefly recall the main properties of the survey. QSO
candidates with optical magnitudes $18.25 \le b_J \le 20.85$ were
selected from the APM catalogue (Irwin, McMahon \& Maddox 1994) in two
$75^{\circ} \times 5^{\circ}$ declination strips centered on
$\delta=-30^\circ$ (South Galactic Cap) and $\delta=0^\circ$ (North
Galactic Cap). The following color selection criteria have been
applied: $(u-b_j)\le 0.36$; $(u-b_j)< 0.12-0.8\;(b_j-r)$; $(b_j-r)<
0.05$ (Smith et al.  2003), in order to guarantee a large photometric
completeness ($ > 90 $ per cent) for quasars within the redshift range
$ 0.3 \le z \le 2.2$.

Spectroscopic observations of the input catalogue were made with the
2-degree Field (2dF) instrument at the Anglo-Australian Telescope. The
spectra were classified both via cross-correlation with specific
templates (AUTOZ, Croom et al. 2004) and by visual inspection.

The final catalogue contains $\sim 21,000$ quasars with reliable
spectra and redshift determinations.  Whenever available, the 2QZ
catalogue also includes radio fluxes at 1.4 GHz from the NRAO VLA Sky
Survey (NVSS; Condon et al. 1998) and X-ray fluxes from the ROSAT All
Sky Survey (RASS; Voges et al. 1999; Voges et al. 2001).

\subsection{Matching Procedure}\label{matching}

In order to determine the radio counterparts of 2QZ quasars we have
matched objects from the FIRST and 2QZ surveys in the equatorial strip
in the North Galactic cap.  In the overlapping region -- $ 9^h \; 50^m
\leq {\rm RA(2000)} \leq 14^h \; 50^m$ and $ -2.8^{\circ} \leq {\rm
Dec(2000)} \leq 2.2^{\circ}$ -- we found 10,110 optical quasars from
the 2QZ.  Over the same area, the total number of sources in the FIRST
survey is $\sim 45,500$ down to a flux limit ${\rm S_{1.4GHz} = 1}$
mJy.  As described in Cirasuolo et al. (2003a; hereafter C03), for the
matching procedure we used an algorithm to collapse multi-component
sources (Magliocchetti et al. 1998): the algorithm collapses
sub-structured sources into single objects having radio fluxes equal
to the sum of the fluxes of the various components. \\ All the
optical-radio pairs having an offset less than 2 arcsec have been
considered as true optical identifications.  The value of 2 arcsec as
matching radius was chosen after a careful analysis as the best
compromise to maximize the number of real associations (estimated to
be $\sim 97$ per cent), and to minimize the contribution from spurious
identifications down to a negligible 5 per cent (Magliocchetti \&
Maddox 2002). Furthermore, in order to verify the reliability of the
associations, all the radio-optical pairs obtained from the collapsing
algorithm were checked by eye on the FIRST image cutouts.

This procedure has lead to 352 quasars with good redshift
determinations from the 2QZ and endowed with radio fluxes ${\rm
S_{1.4GHz} \ge 1}$ mJy over an effective area of 284 square
degrees. Of these sources, 113 have been already presented in C03,
while the remaining 239 constitute an entirely new sample.  The
overall sample will be referred to as FIRST-2dF.

As in C03, the reliability of the present sample has also been checked
against quasars with a radio counterpart in the NVSS (Condon et
al. 1998). The high resolution of FIRST (beaming size 5.4 arcsec)
might in fact lead to a systematic underestimate of the real flux
density in the case of extended sources.  In order to assess the
possible presence and entity of this effect, we have then compared
FIRST and NVSS (beaming size of 45~arcsec) fluxes for all those
sources in the sample which show a radio counterpart in both surveys.
In the common 2QZ, FIRST and NVSS region, we only found 253 objects
having an NVSS counterpart. All of these sources were also detected by
our matching procedure and, as in C03, we found an excellent agreement
between fluxes as measured by FIRST and NVSS. No correction to the
flux densities derived from the FIRST survey was therefore applied.
This also supported the validity of the collapsing algorithm.

About 40 per cent of the sources undetected by NVSS have ${\rm
S_{1.4GHz} \ge 3}$ mJy, i.e. above the flux limit of the survey ($\sim
3$ mJy). This implies that such objects could have been lost in NVSS
because of their multiple component structure.

\subsection{Radio spectral index}\label{radioindex}
Radio spectral indices $\alpha_{\rm R}$
\footnote{Throughout this work the radio flux density is defined as
${\rm S_{\nu} \propto \nu^{-\alpha}}$.} for a sub-sample of FIRST-2dF
sources have been obtained by cross-correlating this dataset with the
Parkes-MIT-NRAO (PMN) radio survey (Griffith et al. 1995).  The PMN
survey covers the equatorial zone ($ -9.5^{\circ} \leq {\rm Dec(2000)}
\leq 10^{\circ}$), observing objects at 4.85 GHz down to a flux limit
of 40 mJy.

\begin{figure}
\center{{
\epsfig{figure=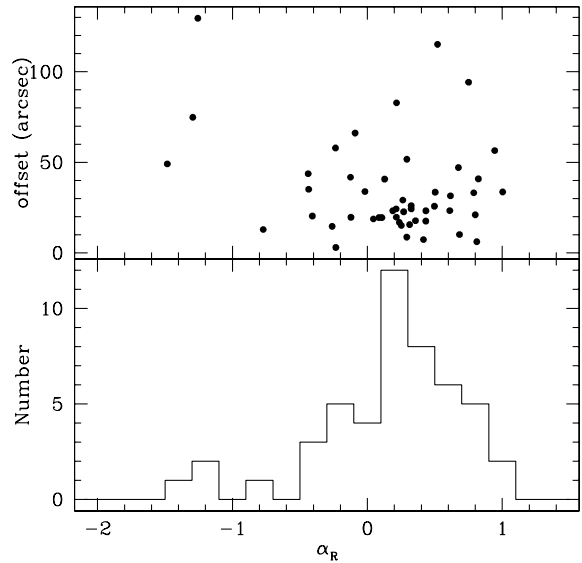,height=8cm}
}}
\caption{\label{alpha} Top panel: radio spectral index for sources in
the FIRST-2dF sample versus the offset between the positions in the
FIRST and PMN surveys.  The lower panel represents the distribution of
the radio spectral indices found.}
\end{figure}

Due to its large beaming size, a matching radius of 3 arcmin has been
used to cross correlate sources in the PMN with those included in our
sample.  Out of the 352 objects, we found counterpart for 52 in the
PMN catalogue. Note that most of the sources indeed present matching
radii $\sim 50$ arcsec, as shown in the upper panel of Figure
\ref{alpha}.  Only a few of them have a larger offset, in any case
smaller than 2 arcmin.  This enables us to be reasonably confident
that they are true associations.

The lower panel in Figure \ref{alpha} shows the distribution of the
radio spectral indices obtained.  It is clear that the majority of the
sources are flat spectrum quasars ($\alpha_{\rm R} < 0.5$). This is
due to the brighter flux limit (${\rm S_{4.85GHz} = 40}$ mJy) of the
PMN survey with respect to FIRST. In fact, only bright steep spectrum
($\alpha_{\rm R} \simgt 0.7$) sources with fluxes ${\rm S_{1.4GHz}
\simgt 100}$~mJy could have been detected by PMN, which therefore
misses the majority of the steep spectrum objects in the FIRST-2dF
sample.  Out of 49 quasars in the FIRST-2dF sample with ${\rm
S_{1.4GHz} \simgt 100}$~mJy we found 13 steep spectrum and 21 flat
spectrum objects.  The remaining 15 sources have been missed in the
PMN survey due to the fact that it does not cover all of the
equatorial region, as it presents holes for $12^h \simlt {\rm RA}
\simlt 16^h$.
%
%

\subsection{The FIRST-2dF sample}
To summarize, the FIRST-2dF sample is comprised of 352 objects with
optical magnitudes $18.25 \le b_J \le 20.85$ and radio fluxes at 1.4
GHz ${\rm S_{1.4 {\rm GHz}}\ge 1}$~mJy. All the 239 objects included
in the FIRST-2dF sample and not included in C03 are presented in Table
2.

\begin{figure}
\center{{ \epsfig{figure=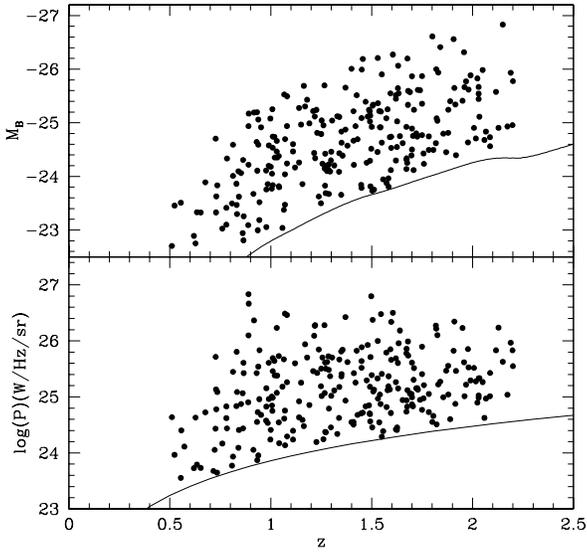,height=8cm} }}
\caption{\label{zpm} Absolute $M_{\rm B}$ magnitude (top panel) and
radio power at 1.4 GHz (lower panel) versus redshift for sources in
FIRST-2dF sample.  The solid lines describe the selection effects
respectively due to the limiting magnitude $b_{\rm J} = 20.85$ and
radio flux density limit ${\rm S_{1.4GHz}} = $ 2 mJy. }
\end{figure}
%

\section{Luminosity Functions}\label{lf}
The larger size of the present sample with respect to the one
presented in C03 (about a factor 3) enables us to obtain the LF of the
population of radio detected quasars, both in the radio and optical
bands.

In order to minimize incompleteness problems, we only considered
sources with ${\rm S_{1.4GHz} > 2}$ mJy (i.e. 95 per cent
completeness).  For the computation of the LFs we further confined the
analysis to the redshift range $0.8 < z < 2.2$.  The upper limit $z
=2.2$ is due to the drop of completeness at higher redshift of the 2dF
QSO Survey (Croom et al. 2004).  The choice of $z=0.8$ as a lower
limit is indeed related to the flux limits of both FIRST and 2dF
surveys.  As shown in Figure \ref{zpm}, faint sources -- either
in the optical or radio bands -- are missed at high redshift due to
selection effects. It follows that, in order to have a complete and
unbiased coverage of both the ${\rm z-M_B}$ and ${\rm z-P_{1.4GHz}}$
planes only sources with $M_B \simlt -24.2$ and \lpow $ \simgt 24.5$
(\whs) can be considered in the analysis (see Figure \ref{zpm}).  In
the redshift range $0.5 \simlt z \simlt 0.8$ the number of sources
satisfying these criteria is extremely small and therefore we
considered only sources with $z \ge 0.8$ in the computation of the
LFs. After the above corrections, the final sample used for the
determination of LFs comprises $\sim 200$ sources.  It is worth noting
that in the following analysis photometric and spectroscopic
incompleteness of the 2QZ survey in the considered redshift range --
both as a function of redshift and apparent magnitude -- has been
taken into account following Croom et al. (2004).

The radio luminosities at {\rm 1.4 GHz} have been calculated by using
- whenever available - the measured $\alpha_{\rm R}$ (see Table 2) or
  adopting a mean value of $\alpha_{\rm R}=0.7$ (see Section \ref{radioindex}).

In order to easily compare our results with those found in the
literature we converted magnitudes from the $b_{\rm J}$ to the B band.
The mean $B-b_{\rm J}$ was computed from the composite quasar spectrum
compiled by Brotherton et al. (2001) from $\sim 600$ radio-selected
quasars in the FIRST Bright Quasar Survey (FBQS).  Actually in the
redshift range $0.3 \le z \le 2.2$ the difference between
corresponding values in the two bands is very small, $0.05 \simlt
B-b_{\rm J} \simlt 0.09$ and independent of redshift. We therefore
chose to apply a mean correction $B= 0.07 + b_{\rm J}$ (see also C03).
The K-correction in the B band has also been computed from the
Brotherton et al. (2001) composite quasar spectrum.

\subsection{Binned ${\rm 1/V_{max}}$ method}\label{vmax}
The simplest approach to estimate LFs and their redshift evolution is
provided by the `classical' ${\rm 1/V_{\rm max}}$ method (Schmidt
1968).  For each source we evaluated the maximum redshift at which it
could have been included in the sample, ${\rm z_{\rm max}= min(z_{\rm
max}^R, z_{\rm max}^O, z_{\rm max}^S)}$, 
where ${\rm z_{\rm max}^R}$ and ${\rm z_{\rm max}^O}$ 
are the corresponding maximum redshifts due to the radio or
the optical limiting flux densities, respectively, and $z_{max}^S=2.2$ 
is the redshift limit of the survey due to completeness.

Radio (RLF) and optical (OLF) luminosity functions have been computed
in four equally spaced redshift bins ($\Delta z = 0.35$) and showed in
the four panels of Figure \ref{lfradio} and \ref{lfopt}, respectively.

\begin{table*}
\begin{center}
\footnotesize
\begin{tabular}{ccccccccc}
\multicolumn{9}{|c|}{Radio LF} \\ \hline
Cosmology & Evol. & $\alpha$   & $\beta$  & ${\rm log_{10}P^*_{1.4}}$ & $k_1$  &  $k_2$ 
	& $\Phi_0$  & $P_{KS}$	\\ \hline \hline
I	& L & $-0.1 \pm 0.1$  &  $1.5 \pm 0.2$ &  $22.7 \pm 0.4$ &  $4.5 \pm 0.5$  &  $-1.5 \pm 0.2$ &  $5.0 \times 10^{-8}$ &  $0.2$ \\
I 	& D & $-0.1 \pm 0.1$  &  $1.6 \pm 0.3$ &  $26.0 \pm 0.2$ &  $2.9 \pm 0.6$  &  $-0.9 \pm 0.2$ &  $2.5 \times 10^{-10}$ &  $0.6$ \\
II	& L & $-0.1 \pm 0.1$  &  $1.5 \pm 0.2$ &  $22.9 \pm 0.5$ &  $4.3 \pm 0.6$  &  $-1.5 \pm 0.2$ &  $4.2 \times 10^{-8}$ &  $0.2$ \\
II 	& D & $-0.1 \pm 0.1$  &  $1.7 \pm 0.4$ &  $26.0 \pm 0.2$ &  $2.9 \pm 0.6$  &  $-0.9 \pm 0.2$ &  $2.0 \times 10^{-10}$ &  $0.5$ \\

\hline

\multicolumn{9}{|c|}{} \\
\multicolumn{9}{|c|}{} \\
\multicolumn{9}{|c|}{Optical LF} \\ \hline
Cosmology & Evol. & $\alpha$   & $\beta$  & $M^*$ & $k_1$  &  $k_2$ 
	& $\Phi_0$  & $P_{KS}$	\\ \hline \hline
I	& L & $0.2 \pm 0.6$  &  $2.4 \pm 0.5$ &  $20.4 \pm 1.0$ &  $2.3 \pm 0.5$  &  $-0.8 \pm 0.2$ &  $6.3 \times 10^{-8}$ &  $0.6$ \\
I 	& D & $0.7 \pm 0.2$  &  $2.9 \pm 0.7$ &  $25.6 \pm 0.4$ &  $1.3 \pm 0.5$  &  $-0.5 \pm 0.1$ &  $5.3 \times 10^{-9}$ &  $0.5$ \\
II	& L & $0.6 \pm 0.3$  &  $2.8 \pm 0.3$ &  $21.1 \pm 1.0$ &  $2.0 \pm 0.5$  &  $-0.6 \pm 0.2$ &  $4.4 \times 10^{-8}$ &  $0.4$ \\
II 	& D & $0.8 \pm 0.2$  &  $2.8 \pm 0.7$ &  $25.6 \pm 0.5$ &  $1.4 \pm 0.5$  &  $-0.5 \pm 0.2$ &  $3.1 \times 10^{-9}$ &  $0.5$ \\

\hline
\end{tabular}
\caption{\label{param} Best fit parameters for the radio and optical
LF of the FIRST-2dF sample. The characters L and D in the second
column refer to a model of luminosity (Eq. \ref{lum_evol}) or density
(Eq. \ref{dens_evol}) evolution, respectively.}
\end{center}
\end{table*}
%

\subsection{Parametric method}\label{like}
To obtain an independent and more quantitative description of the LFs
and to avoid loss of information due to the binning process, we also
carried out a maximum likelihood analysis.  This is a parametric
technique which relies on maximizing the probability of observing
exactly one quasar in a $\Delta z \Delta L$ element at each redshift
and luminosity for all the quasars in the data set and of observing
zero objects in all the other differential elements in the accessible
regions of the redshift-luminosity plane (Marshall et al. 1983).  This
method requires an analytic functional form for the LF
and its evolution.

We chose to model both the radio and optical LFs as double power laws
in luminosity (or magnitude) as:
\begin{equation}
\Phi(P_{1.4},z) = \frac{\Phi_0}{(P_{1.4}/P_{1.4}^*)^\alpha + 
      (P_{1.4}/P_{1.4}^*)^\beta}
\end{equation}

and 
\begin{equation}
\Phi(M_B,z) = \frac{\Phi_0}{10^{0.4(1-\alpha)(M_B - M^*_B)} +
10^{0.4(1-\beta)(M_B - M^*_B)}}.
\end{equation}
The time evolution of the LFs has been modeled with a 2nd-order
polynomial function both in the case of pure luminosity evolution:
\begin{equation}\label{lum_evol}
P_{1.4}^*(z) = P_{1.4}^*(z=0) \;\times \: 10^{k_1 z + k_2 z^2},
\end{equation}
and
\begin{equation}
M^*_B(z) = M^*_B(z=0) -2.5 (k_1 z + k_2 z^2)
\end{equation}
where $P_{1.4}^*$  and $M^*_B$ are the radio and optical break luminosity,
respetively,  and pure density  evolution:
\begin{equation}\label{dens_evol}
\Phi_0(z) = \Phi_0(z=0) \; \times \: 10^{k_1 z + k_2 z^2}.
\end{equation}

\begin{figure*}
\center{{ \epsfig{figure=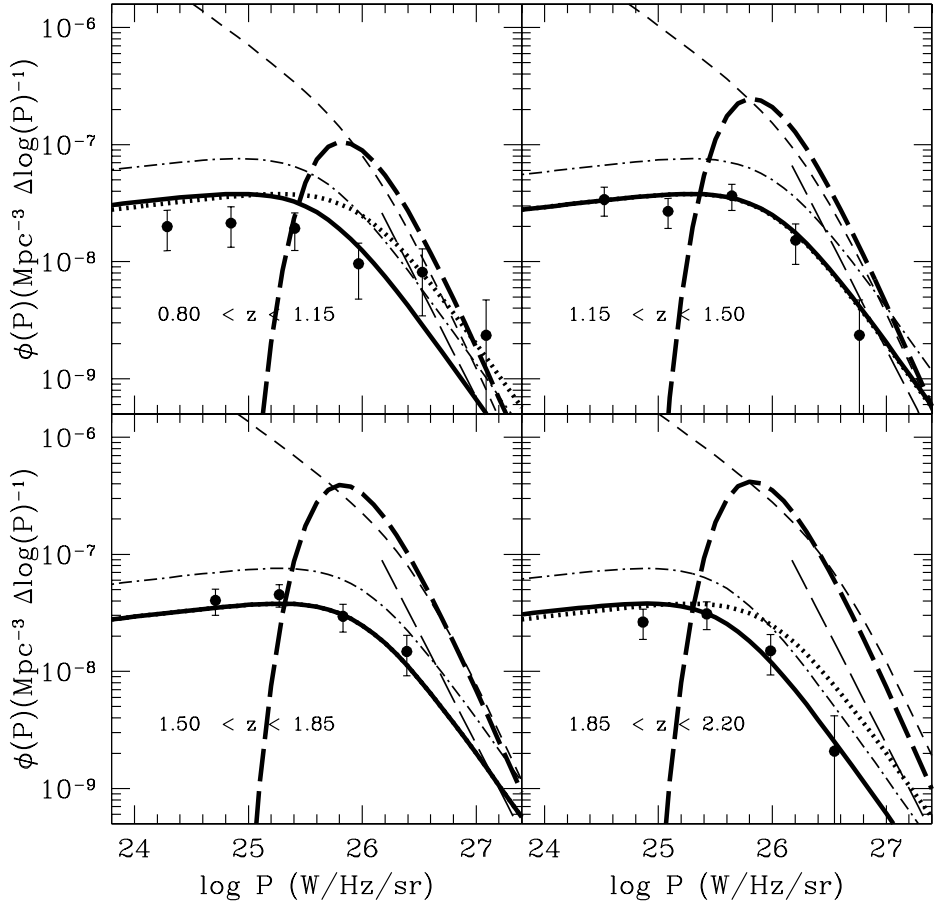,height=13cm} }}
\caption{\label{lfradio} Radio luminosity functions for sources in the
FIRST-2dF sample. The filled circles represent the LF obtained with
the $\rm{1/V_{\rm max}}$ method and the solid lines those derived from
the likelihood analysis, by assuming a luminosity evolution model in
cosmology I. The dotted lines indicate the RLF at the redshift of its
maximum, while the dot-dashed lines are the RLFs corrected in
normalization by a factor of 2 (see text). For comparison, we also
plotted the LFs from Dunlop \& Peacock (1990) (dashed), Willott et
al. (1998) for steep spectrum radio loud quasars (thin long-dashed
lines) and Willott et al. (2001) for the high luminosity radio
population (thick long-dashed lines).}
\end{figure*}

The best fit parameters have been obtained by minimizing the
likelihood function, through the MINUIT package from the CERN
libraries.  The values from the fitting are given in 
Table \ref{param} for both the
radio and optical LFs. Errors on each parameter, corresponding to 68
per cent confidence level, have been obtained by using the MINOS task
in MINUIT.  The algorithm procedure consists in varying one parameter
at a time, and then minimizing the likelihood function with respect to
all the other parameters, in order to find the variation in the
function corresponding to one standard deviation error. Figures
\ref{lfradio} and \ref{lfopt} show the best-fit solutions obtained
from the likelihood analysis (solid curves), compared with the LFs
obtained with the binned method (Section \ref{vmax}). The analytic
functions have been computed at the mean redshift of each bin.  The
agreement between the LFs obtained through these two independent
methods is remarkably good.

To have a quantitative assessment of the goodness of the best-fit
parameters we used a 2D Kolmogorov-Smirnov (KS).  This
multidimensional version of the KS test (Press et al. 1992) compares
the observed redshift-luminosity plane with the one obtained
integrating the LF. Note that the KS test can only be used to reject
models when the resulting probability $P_{\rm KS}$ is less than 0.1 or
0.05.  In our case the probability $P_{\rm KS}$ values are $\ge 0.2$
(see Table \ref{param}), implying that the data and model are not
significantly different.

Note that in the considered redshift range ($0.8 \simlt z \simlt
  2.2$) the radio LFs obtained with the two evolution models are very
  similar, since we are sampling only the faint end of the LF. Figures
  \ref{lfradio} and \ref{lfopt} show the results for the luminosity
  evolution model. In order to provide a more direct visualization
of the redshift evolution of the LFs, in each panel of Figures
\ref{lfradio} and \ref{lfopt} we also reported -- as a dotted line --
the LF at the maximum of its redshift evolution which is reached in
both the radio and optical bands at $z \sim 1.7 \pm 0.2$.  We note
that this preliminary analysis suggests the presence of a decline in
the observed number density of objects at $z \sim 2$.  Even though
this decline is not statistically significant and more data are needed
to verify the trend, this appears to hold in both bands. We will
present stronger evidence for such a finding in Section \ref{result}.

%
%
\subsection{Assessment of incompleteness}\label{incompl}
It is worth noting that the LFs presented in the previous Section have
been computed by applying various cuts to the dataset, both in
redshift and luminosity.  In fact, the analysis has been confined to
a particular redshift range ($0.8 \le z \le 2.2$). Moreover, limits
have been applied to both radio and optical luminosities, in order to
avoid incompleteness effects and obtain an unbiased coverage of the
redshift-luminosity planes.  In this section we try to estimate the
effects of these limits on the derived LFs, with particular attention
devoted to the $M_B \simlt -24.2$ and \lpow $ \simgt 24.5$ (\whs)
cuts.

In order to provide an estimate of the incompleteness affecting the
radio LF, we considered the redshift interval $0.5 \simlt z \simlt
1.2$, in which the coverage of the ${\rm z-M_B}$ plane is complete
down to $M_B \sim -23$ (see Figure \ref{zpm}).  In this redshift
range, we computed the RLF applying either a cut at $M_B \le -24.2$ or
$M_B \le -23$. It turns out that the shape of the RLF does not change
by applying these two limits, while the normalization obtained in the
case of $M_B \le -24.2$ is about $2$ times lower than that found in
the $M_B \le -23$ case.  We stress that due to the small statistics in
this redshift range such result should only be taken as a rough
estimate of the true incompleteness.  Therefore, under the ``strong''
assumption that the $M_B \le -24.2$ cut only affects the normalization
of the RLF, independently of redshift, we can correct the RLF by a
factor 2 in normalization, in order to have a rough estimate of the
``complete'' RLF of quasars down to $M_B \le -23$.
  
We also tested for the effects of the \lpow $\ge 24.5$ (\whs) cut on
the OLF. In the same redshift range ($0.5 \simlt z \simlt 1.2$) we can
achieve completeness down to \lpow $\ge 24$ (see bottom panel of
Figure \ref{zpm}).  An analysis analogous to that applied for the RLF
reveals that the cut at \lpow $\ge 24.5$ (\whs) does not affect the
shape of OLF, as in the case of the RLF, but only its normalization by
a factor $\sim 1.5$.  Both radio and optical LFs corrected for this
incompleteness, assumed to hold at all redshifts (i.e. for LFs with a
shape independent of $z$), are shown as dot-dashed lines in Figures
\ref{lfradio} and \ref{lfopt}.

\section{Comparison with previous results}\label{result}

\subsection{Radio Luminosity Function}
Our results on the RLF can be meaningfully compared at its bright end
with those obtained for bright radio selected quasars by Willott et
al. (1998).  A sample of steep spectrum quasars with $M_B < -23$ was
selected by these authors from the 7C (McGilchrist et al. 1990) and
3CRR (Laing et al. 1983) catalogues.  All their sources are found to
be radio loud, due to the bright limits of these low frequency
surveys, ${\rm S_{151MHz} \ge 0.51}$ Jy and ${\rm S_{178MHz} \ge
10.9}$ Jy, respectively.

Before moving on, a caveat is necessary. Our dataset cannot
discriminate between steep and flat objects.  As discussed in Section
\ref{radioindex}, most of the sources in the FIRST-2dF sample are
probably steep spectrum, as suggested by the lack of detection in the
PMN. Moreover, at low radio powers (\lpow $\simlt 26$ \whs), the
number of flat spectrum sources is much smaller than the steep
spectrum ones even in samples selected at high frequencies (Dunlop \&
Peacock 1990; Maraschi \& Rovetti 1994).  Therefore, this allows us to
directly compare our data with the steep spectrum dataset of Willott
et al. (1998).

In order to perform this comparison, we have converted source
luminosities from the Willott et al. (1998) sample to 1.4 GHz by
assuming a radio spectral index $\alpha_{\rm R}=0.7$.  The RLF for
their best fit (model C) are shown in Figure \ref{lfradio} as thin
long dashed lines.

These appear to be in reasonably good agreement with our RLF
(corrected for renormalization, see Section 3.3).  In fact, the slopes
of the bright ends are similar, even though the statistics of our data
is very poor in this luminosity range.  Our best fit value for the
slope is $1.7 \pm 0.3$, fully consistent with $1.9 \pm 0.1$ quoted by
Willott et al. (1998).  The second important aspect is related to the
flattening of the RLF at the faint end. Due to the paucity of data,
Willott et al.  estimated the value of the break luminosity by fixing
a slope $\alpha = 0$ for the faint end and fitting the radio number
counts at $0.1$ Jy.  This procedure led to a value of the break
luminosity ${\rm log_{10}P^*_{151MHz} \sim 26.8}$ (\whs), i.e.  $\sim
26.0$ when translated at 1.4 GHz.  With our sample -- which is more
than 100 times fainter than the 7C -- we can directly determine for
the first time this flattening and our analysis confirms that the
faint end is well described by $\alpha \sim 0$. Furthermore, we
infer a break at a luminosity \lpow $\sim 26$ \whs in agreement with
that deduced by Willott et al. (1998), by considering both a
luminosity and a density evolution model.

The evolution of the RLF found in our sample is also consistent with
that derived for brighter sources, showing a maximum at $z \sim
1.7$. However, the FIRST-2dF sample shows indications for negative
evolution beyond $z \sim 1.8$ (see Section \ref{like}), while the
brighter Willott et al. sample is consistent with no evolution up to
the highest redshifts probed by their surveys ($z \sim 3$). However,
their statistics in the $2 \simlt z \simlt 3$ range is rather limited
($< 10$ sources).  

In order to assess the significance of our finding of a negative
evolution beyond $z \sim 1.8$, we considered the behaviour of the
quasar number density $\rho(z)$ as a function of redshift.  This is
obtained by integrating the LFs over the luminosity range detectable
up to the highest redshift in our data.  It is worth noting that, even
though we do not have information on the bright end of the RLF, our
sample extends up to the break of the LF, thus accounting for the bulk
of the sources.  The number density $\rho(z)$ is plotted in Figure
\ref{density}, where the error bars indicate Poissonian uncertainties.
We find that the space density of FIRST-2dF quasars shows a
significant ($\sim 3 \sigma$) tendency for a decrement at $z \simgt
1.8$. Such a decrement is found to be approximately a factor 2 at
$z=2.2$.   These results hold in both cosmologies.  In the small
panel of Figure \ref{density} we also show the space density computed
for two bins of radio power. The decline observed for the whole sample
is still evident for low-luminosity objects as well as for brighter
ones.  For the latter sources the decline seems to be less pronounced,
but the small statistics does not allow to draw any conclusion.

We note that our choice of adopting a single spectral index
($\alpha_{\rm R} = 0.7$) for all the sources undetected by PMN could
introduce some uncertainties in the determination of $\rho(z)$.  In
fact, radio spectra steepen at higher frequencies and, as shown by
Jarvis \& Rawlings (2000), this is crucial in determining the space
density of radio objects at high redshift.  In order to quantify this
uncertainty we have then considered an ``extreme case'' by assuming
for all the sources with $z>1.7$ a steep spectral index ($\alpha_{\rm
R}=1.5$).  Clearly this results in larger radio powers and in turn in
a rise in the bright end of the LF, to the detriment at fainter
luminosities.  The crucial point is to assess the completeness of the
sample. In fact, due to the steeper spectral index at high z, the
observed fluxes would drop more rapidly with redshift. However, even
by assuming such a steep spectral index ($\alpha_{\rm R} =1.5$) for
$z>1.7$, the 2 mJy limit of our sample ensures completeness down to
\lpow $\sim 25$ \whs for $z \le 2.2$.  This, together with the fact
that the trend for the space density of fainter and brighter sources
is found to be similar (see Figure 4), makes us confident that our
results, and particularly those on the space density of radio-loud
quasars, are not severely affected by the uncertainty associated to an
unknown spectral index.

Another source of uncertainty could arise from the double selection,
radio and optical, of the sample since radio and optical luminosities
are related, even though with a large scatter (Cirasuolo et
al. 2003b). Therefore, low luminosity radio sources would be affected
by the optical flux limit more quickly than the brighter ones. This
selection effect would be effective at all redshifts, but stronger at
high z due to the redshift dependence of the limiting luminosity (see
Figure \ref{zpm}), producing a fake decline in the space density.
However, the large scatter between radio and optical luminosities --
at least three orders of magnitude in radio luminosity at a fixed
optical one (Cirasuolo et al. 2003a) -- completely dominates over this
selection effect.  This is again supported by the similar decline in
space density at high redshift of sources with different radio powers
(see Figure \ref{density}).

The last point worth stressing is that we find no evidence for a
strong decrement in the space density at low redshifts.  In fact,
$\rho(z)$ is found to decrease by only a factor $\sim 1.5$ between $z
\sim 1$ and $z \sim 0.5$.  This lack of a significant evolution at low
redshifts has also been recently claimed by Clewley \& Jarvis (2004)
for fainter sources (${\rm P_{325MHz} < 10^{25}}$ \whs) at $z < 0.5$,
even though the evolution of the very low luminosity population
(i.e. FR~I) still remains unconstrained (Magliocchetti, Celotti \&
Danese 2002). 

\subsection{Optical Luminosity Function}
Clearly the best reference for the OLF of the radio quasar sample is
constituted by the OLF of the `parent' (2dF QSO Redshift Survey)
population (Croom et al. 2004).

\begin{figure*}
\center{{
\epsfig{figure=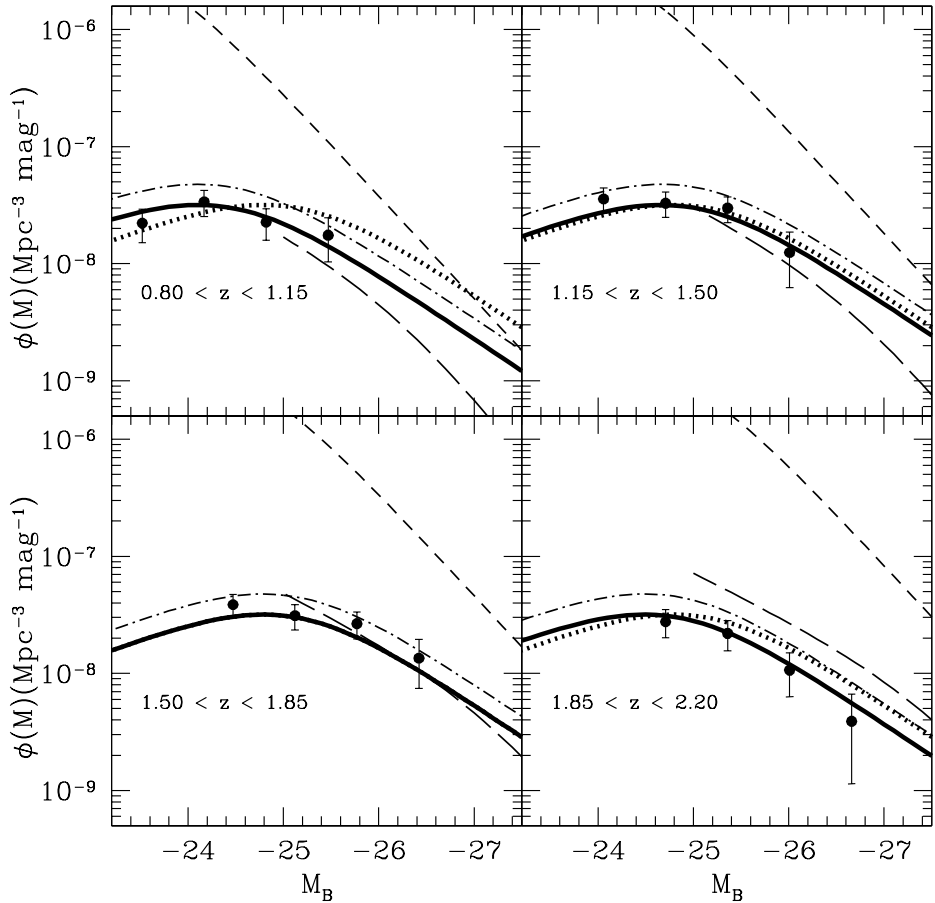,height=13cm}
}}
\caption{\label{lfopt} Optical luminosity functions for sources in the
FIRST-2dF sample. The filled circles represent the LF obtained with
the $\rm{1/V_{\rm max}}$ method and the solid lines those derived from
the likelihood analysis, by assuming a luminosity evolution model in
cosmology I. The dotted lines indicate the OLF at the redshift of its
maximum, while the dot-dashed lines are the OLFs corrected in
normalization by a factor of 1.5 (see text). For comparison, we also
plotted the LFs from La Franca et al. (1994) (long-dashed) and for the
quasar population as a whole (Croom et al. 2004, dashed lines).}
\end{figure*}

\begin{figure}
\center{{ \epsfig{figure=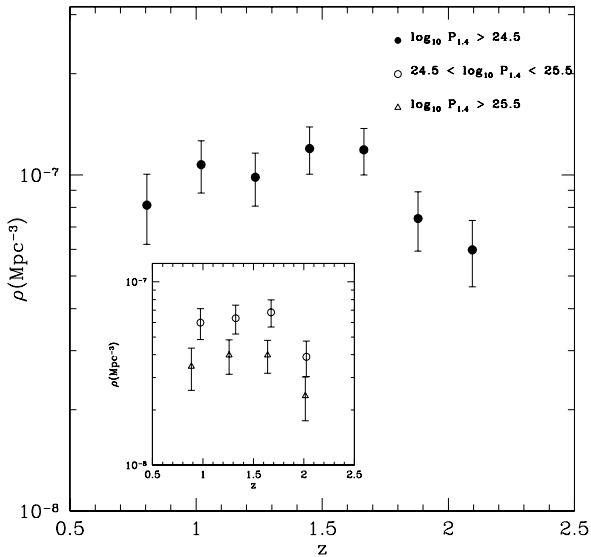,height=8cm} }}
\caption{\label{density} Space density as a function of redshift for
sources in the FIRST-2dF sample for cosmology I. The small panel shows
the same space density for two different bins in radio power.}
\end{figure}

In Figure \ref{lfopt} we report the best fit solutions for the OLF
found by Croom et al. (2004) (dashed lines, corresponding to the
polynomial evolution model in the case of cosmology I).  As in the
case of the RLF, we applied a correction of a factor $1.5$ in the
normalization of our OLF to account for incompleteness effects (see
Section \ref{incompl}). This provides an estimate of the OLF
(dot-dashed lines) representative of the whole population of
radio-active quasars with \lpow $\ge 24$ (\whs).

As expected, the OLF of radio detected quasars is flatter than the one
derived for the population as a whole. This is already visible at the
brighter end where the slope of the FIRST-2dF OLF is $\beta \sim 2.4 -
2.8$ (depending on the adopted model), while that obtained by Croom et
al. (2004) is $\beta = 3.2$. This trend becomes increasingly more
significant towards fainter magnitudes. Indeed, this is consistent
with the observed dependence of the fraction of radio detected quasars
as a function of optical luminosity (Padovani 1993; La Franca et
al. 1994; Hooper et al. 1995; C03).  However, Cirasuolo et al. (2003b)
showed that this can simply be the result of selection effects due to
the radio and optical limits of the various samples, under the
assumption that the optical and radio luminosities are linearly (even
though) broadly correlated.  In this case the intrinsic shape of the
distribution of the radio-to-optical ratio $R^*_{1.4 GHz}$ for the
quasar population as a whole is found to show a broad peak at
$R^*_{1.4 GHz}\sim 0.3 $ containing more than 90 per cent of the
sources, followed by a steep transition region at $1 \simlt
R^*_{1.4GHz} \simlt 10$ (see Fig. 4 in Cirasuolo et al. 2003b). This
implies that, at a given radio flux limit, surveys with brighter
optical limiting magnitudes preferentially select smaller values of
$R^*_{1.4 GHz}$ -- i.e. closer to the peak of the distribution --
which results in a higher percentage of radio detections.

In Figure \ref{lfopt} we also compare our results with previous
determinations of the OLF for bright ($M_B \simlt - 25$) radio-loud
quasars (La Franca et al. 1994; their best-fit function (model B) is
indicated with long-dashed lines).  An overall good agreement between
the two OLFs is found in the magnitude range in which the two samples
overlap.

The OLF obtained from our sample peaks at $z \sim 1.7$ and -- as in
the case of the RLF -- possibly hints to a decline at higher
redshifts. On the contrary, La Franca et al. (1994) find a maximum
redshift for luminosity evolution $\rm{z_{max} \sim 1.9}$, beyond
which no further evolution is detected.  The same substantial lack of
evolution in the range $1.5 \simlt z \simlt 2$ has also been claimed
by Goldschmidt et al. (1999).  However, we argue that the limited
sizes of their $z\sim 2$ samples do not allow for any strong
statistical conclusion.

\subsection{The whole radio source population}
\begin{figure}
\center{{ \epsfig{figure=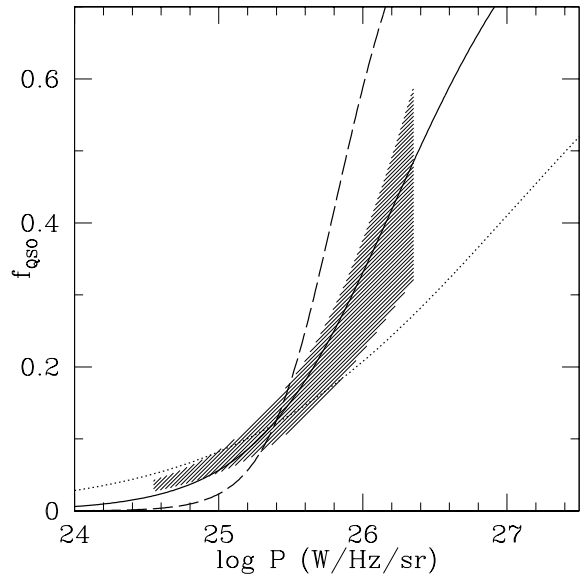,height=8cm} }}
\caption{\label{torus} Fraction ${\rm f_{QSO}}$ of radio quasars from
the FIRST-2dF sample with respect to the whole radio population (taken
from Dunlop \& Peacock 1990) as a function of radio power (shaded
area). The solid line is the predicted fraction on the basis of
equation \ref{eqtorus} assuming $\delta=1$, while the dotted and
dashed lines respectively correspond to $\delta=0.5$ and
$\delta=2$. The predicted fractions have been normalized at \lpow
=25.5 \whs. }
\end{figure}

Let us finally consider the relation between the radio quasar and the
entire radio source population. The dashed lines in Figure
\ref{lfradio} represent the RLF for steep plus flat spectrum radio
sources obtained by Dunlop \& Peacock (1990) (calculated at 1.4 GHz by
assuming a radio spectral index $\alpha_{\rm R}=0.7$ for steep
spectrum sources). It is clear that the fraction of radio quasars
decreases toward lower powers, implying that for \lpow $\simlt 26$
(\whs) the fraction of radio galaxies is dominant with respect to
radio quasars, while at higher luminosities the two populations are of
comparable importance.  The shaded area in Figure \ref{torus} shows
the ratio $f_{\rm QSO}$ between the RLF of radio quasars -- drawn from
the FIRST-2dF sample and corrected in normalization -- and the RLF of
Dunlop \& Peacock (1990) as a function of radio luminosity.  The
increasingly larger scatter for \lpow $\simgt 25.5$ (\whs) is due to
the different evolution in redshift of the two RLFs.  As already
discussed, the space density of radio active quasars is found to
decline at high redshift, while the radio population as a whole still
shows a positive evolution, therefore increasing the scatter in
$f_{\rm QSO}$.  A similar trend in the fraction of quasars as a
function of luminosity has been found by Willott et al. (2000) by
analyzing radio selected samples.  They found that this fraction is
$\sim 40$ per cent at bright (narrow [OII]) emission line luminosities
($L_{[OII]}$) -- assumed to be good tracer of the ionizing continuum
-- dropping to a few per cent for fainter luminosities.  They
suggested that this finding can be accounted for by either a gradual
decrease of the opening angle of the obscuring `torus' with decreasing
ionizing luminosity or with the emergence of a distinct population of
low luminosity radio sources, M87--like.

Following the approach suggested by Willott et al. (2000) we try to
interpret the decrease of $f_{\rm QSO}$ within the `receding torus'
scenario. In this model, initially proposed by Lawrence (1991; see
also Simpson 1998), the inner radius of the torus depends on the AGN
radiation field which sublimates the dust grains. If the scale-height
of the torus is independent of the ionizing luminosity $L_{\rm ion}$,
its inner radius scales as $L_{\rm ion}^{0.5}$ resulting in larger
half-opening angles $\theta$ at higher luminosities.  The fraction of
quasars predicted is then:
\begin{equation}\label{eqwill} 
{\rm f_{QSO}} = 1 - \left( 1 + \left( \frac{L_{\rm ion}}{L_0} \right)
\tan^2 \theta_0 \right)^{-0.5},
\end{equation}
where $L_0$ and $\theta_0$ are a normalization luminosity and the
corresponding torus half-opening angle, fixed by the measured quasar
fraction at that luminosity.  As we cannot rely on the $L_{[OII]}$ as
Willott et al., as a working hypothesis we use the radio luminosity as
a tracer of the central ionizing radiation field. This is based on the
linear relation between radio power at 151 MHz and $L_{[OII]}$ suggested by
Willott (2000), $L_{[OII]} \propto L_{151}^{1.0}$, although
with a large scatter (see also Willott et al. 1999). 
Because of this uncertainty, we modified
Eq. \ref{eqwill} assuming that the ionizing luminosity scales as
$P^{\delta}$, where $P$ is the radio power at 1.4 GHz and $\delta$ is
a free parameter, i.e.:
\begin{equation}\label{eqtorus} 
{\rm f_{QSO}} = 1 - \left( 1 + \left( \frac{P}{P_0} \right) ^{\delta}
\tan^2 \theta_0 \right)^{-0.5}.
\end{equation}
Different values for the parameter $\delta$ have been tested in order
to reproduce the measured fraction of radio active quasars. As shown
in Figure \ref{torus}, a nearly linear dependence of the central
ionizing luminosity on the radio power ($\delta \sim 1$) is able to
reproduce the observed fraction, $\delta\simeq 0.5$ is only marginally
consistent with it, while the extreme value $\delta=2$ is totally
inconsistent with the observed behaviour. This finding is a posteriori
broadly consistent with the relation $L_{[OII]}$ vs $L_{151}$ found by
Willott (2000) and with the evidence found by Cirasuolo et
al. (2003b) for a nearly linear dependence of the radio luminosity on
the optical one.

However, the linear relation used in Eq. (\ref{eqtorus}) does not take into
account the large scatter between radio and optical/UV luminosities, which 
could mask the effect postulated by the receding torus model.  
An observational way to test whether indeed $\simgt 90$\% of weak 
radio  sources
contain a hidden quasar -- as implied by such model --  
would be the measure of the level of their nuclear (disc/torus) activity 
through far-IR photometry (i.e. with SPITZER).

The comparison between the radio quasar and the whole radio source
population within the framework of the unification scenario --
postulating that radio-loud quasars represent the beamed counterparts
of powerful (FR~II) radio sources -- reveals a further aspect worth
mentioning.  Willott et al. (2001) proposed a dual population model
for the RLF by separating the contribution from the high and low
luminosity sources assumed to have a differential density
evolution. The latter population can be associated with weak/low
ionization emission lines radio galaxies -- both FR~I and FR~II --
while the high luminosity population can be representative of radio
galaxies and quasars with strong/high ionization emission lines and
FR~II radio properties.  A comparison of our RLF with that of the high
power population identified by Willott et al. (2001) (converted to 1.4
GHz by assuming $\alpha_{\rm R}=0.7$; thick long-dashed lines in
Figure \ref{lfradio}) shows that at \lpow $< 25$ (W Hz$^{-1}$
sr$^{-1}$) the observed quasar number density largely exceeds that of
the whole population \footnote{Formally, the low luminosity end of the
distribution of the Willott et al. (2001) powerful population actually
cuts even above the definition of FRII-like sources (\lpow $\sim 24.5$
\whs at 1.4 GHz).} as derived by Willott et al.

It is worth reminding here that all of the sources in the FIRST-2dF
sample are classified as quasars, based on both their broad emission
lines and bright luminosity ($M_B \le -23$). Furthermore, they are
radio loud according both to their radio luminosity (\lpow $\ge 24$
\whs; Miller et al. 1990) and radio-to-optical ratio ($R_{1.4 GHz}^*
\ge 30$; Kellermann et al. 1989).  However, for the first time, the
FIRST-2dF sample is able to explore the low optical and radio
luminosity ranges. In particular, the radio powers sampled with this
work approach the transition region between FRII and FRI sources,
posing the problem of the nature of these low luminosity objects.

A few possibilities are open.  They could hint at a population of
quasars with intermediate FRII-FRI radio properties, either intrinsic
or due to an evolution between these two radio phases (such as those
discussed by Blundell \& Rawlings 2001).  Alternatively, these objects
might be bordering the intermediate radio quasar population (i.e.
between loud and quiet radio quasars with no evidence for relativistic
jets). Finally,  
they might simply represent a lower luminosity version
of powerful radio quasars, which thus do not obviously
fit into the `classical' scheme for radio sources. This might
correspond to the fading phase of luminous quasars. Interesting clues
on this issue would come from the study of their radio morphology.

\section{Discussion}
The novelty of our work is that, by exploiting the faint observational
limits of both radio and optical surveys such as the FIRST and the 2dF
Quasar Redshift Survey, it was possible for the first time to explore
the faint end of the LFs of radio (loud) quasars at both wavelengths.
Furthermore, the larger number of radio-detected quasars collected by
using these datasets allowed for a
suitable determination of these LFs up to $z = 2.2$.

The most intriguing result we found is the indication ($\sim 3
\sigma$) -- both in radio and optical band -- of a negative evolution
for these faint sources at $z> 1.8$.  This corresponds to a decrement
in the space density of faint quasars of about a factor 2 at $z\simeq$
2.2.  For brighter -- either radio or optically selected -- samples,
the evolution of radio--loud quasars shows a peak at $z\sim 1.7 -
2.0$, but no evidence for a decline at higher redshifts (Willott et
al. 1998; La Franca et al. 1994; Goldschmidt et al. 1999).  This
result is thus consistent with the presence of a differential
evolution for the population of radio-active quasars (Vigotti et
al. 2003), where the weakest radio sources show a more pronounced
decline in space density than the more powerful ones, suggesting the
evolution to be a function of the intrinsic power (as in the
luminosity/density evolution model of Dunlop \& Peacock 1990).

The second interesting finding is the flattening of the faint end of
the LFs of radio active quasars in both the radio and optical bands.
The found behaviour of the OLF at faint magnitudes is consistent with
previous estimates performed at brighter magnitudes (La Franca et al.
1994; Goldschmidt et al. 1999), and the comparison with the OLF of the
quasar population as a whole supports a dependence of the fraction of
radio detected quasars on their optical luminosity.  This flattening
can simply result from a linear although broad intrinsic correlation
between the radio and optical luminosities when combined with the
radio and optical limits of the various surveys (Cirasuolo et
al. 2003b).

Finally, we find that a progressive decrement in the fraction of
quasars in the whole radio source population can be consistently
accounted for within the `receding torus' scenario by assuming a
quasi-linear dependence between optical and radio luminosities. At
this stage, the nature of the lower radio luminosity sources in our
sample appears unclear substantially due to the lack of information on
the origin/characteristics of their radio emission, which could be
unveiled by higher resolution radio imaging.

\noindent 
\section{\bf Acknowledgments} 

We wish to thank Luigi Danese for insightful discussions and the
referee for helpful comments. We also acknowledge J.A. Peacock and G. 
Zamorani for useful suggestions. The Italian MIUR and INAF are
acknowledged for financial support.


\end{document}